\documentclass{llncs}

\usepackage{diagrams}

\usepackage{oldlfont,amssymb,epsf,stmaryrd}
    
\newbox\tempa
\newbox\tempb
\newdimen\tempc
\def\mud#1{\hfil $\displaystyle{\mathstrut #1}$\hfil}
\def\rig#1{\hfil $\displaystyle{#1}$}
\def\irulehelp#1#2#3{\setbox\tempa=\hbox{$\displaystyle{\mathstrut #2}$}%
		        \setbox\tempb=\vbox{\halign{##\cr
	\mud{#1}\cr
	\noalign{\vskip\the\lineskip}%
	\noalign{\hrule height 0pt}%
	\rig{\vbox to 0pt{\vss\hbox to 0pt{${\; #3}$\hss}\vss}}\cr
	\noalign{\hrule}%
	\noalign{\vskip\the\lineskip}%
	\mud{\copy\tempa}\cr}}%
		      \tempc=\wd\tempb
		      \advance\tempc by \wd\tempa
		      \divide\tempc by 2 }
\def\irule#1#2#3{{\irulehelp{#1}{#2}{#3}%
		     \hbox to \wd\tempa{\hss \box\tempb \hss}}}

\newcommand{\fa}{\forall}
\newcommand{\ex}{\exists}
\newcommand{\ra}{\rightarrow}

\newcommand{\barity}[1]{\langle #1 \rangle}

\newcommand{\VC}[1]{} \newcommand{\VL}[1]{#1}

\begin{document}

\title{Binding Logic: proofs and models}

\author{Gilles Dowek\inst{1}
\and
Th\'er\`ese Hardin\inst{2}
\and
Claude Kirchner\inst{3}}

\institute{
INRIA-Rocquencourt, B.P. 105, 78153 Le Chesnay Cedex, France,\\
{\tt Gilles.Dowek@inria.fr}, {\tt http://logical.inria.fr/\~{}dowek} 
\and
LIP6, UPMC, 8 Rue du Capitaine Scott, 75015
Paris Cedex, France,\\
{\tt Therese.Hardin@lip6.fr},
{\tt http://www-spi.lip6.fr/\~{}hardin}
\and
LORIA \& INRIA,
615, rue du Jardin Botanique, 
54600 Villers-l\`es-Nancy, France\\
{\tt Claude.Kirchner@loria.fr},
{\tt http://www.loria.fr/\~{}ckirchne}}

\date{}

\maketitle

\thispagestyle{empty}

\begin{abstract}
We define an extension of predicate logic, called \textit{Binding
Logic}, where variables can be bound in terms and in propositions. We
introduce a notion of model for this logic and prove a soundness and
completeness theorem for it. This theorem is obtained by encoding this
logic back into predicate logic and using the classical soundness and
completeness theorem there.
\end{abstract}

In predicate logic, only a quantifier can bind a variable.  Function
and predicate symbols cannot.  In contrast, in the informal language of
mathematics and in languages used in computer science, binders are
pervasive, as they are used to express functions, relations, definite
descriptions, fix-points, sums, products, integrals, derivatives, etc.

There are many ways to express such constructions without explicit
binding symbols using, for instance, combinators \cite{Curry}, de
Bruijn indices \cite{dB}, higher-order abstract syntax
\cite{MillerNadathur,PfenningElliott}, \ldots However, as binders are
everywhere in mathematics and in computer science, we believe that
they should be studied as a basic notion, and not just encoded.  The
first contribution of this paper is to define a logic with binders,
called {\em Binding Logic}, that is a simple extension of predicate
logic: proof rules are the same and no extra axiom is added.

A first question, about binding logic, is that of the provability of
the extensionality scheme
$$(\fa x~(t = u)) \Rightarrow \Lambda x~t = \Lambda x~u$$ 
where $\Lambda$ is a unary function symbol binding one
variable in its argument. 
This scheme cannot be derived in a simple way from the axiom of equality
\cite{Hindley}:
$$\fa y~\fa z~(y = z \Rightarrow \Lambda x~y = \Lambda x~z)$$ because
substituting $t$ for $y$ and $u$ for $z$ renames the variable $x$ to
avoid captures.  We shall prove that this scheme is indeed
independent.  Thus, if we want binders to be extensional, we must add
this scheme, but Binding logic can also be used, if we do not, for
instance, when we define weak reduction in $\lambda$-calculus or when
we want to distinguish various algorithms that implement the same
function.

To prove such independence results, we want to construct
counter-models. Thus the notion of model of predicate logics needs to
be extended to Binding logic.  The last and main contribution of this
paper is to define such a notion of model, and prove that is sound and
complete with respect to deduction.

In predicate logic, a term $t$ that contains a free variable $x$ does
not have a unique denotation in a given model ${\cal M}$. Instead, for
each element $a$ of the domain $M$ of this model, we can define a
denotation $\llbracket t \rrbracket_{a/x}$. It is tempting to define
the absolute denotation of the term $t$ as the function from $M$ to
$M$ mapping $a$ to $\llbracket t \rrbracket_{a/x}$.  Then, we would
obtain the denotation of $\Lambda x~t$ by applying the denotation of
the symbol $\Lambda$ to the function denoted by the term $t$. The
denotation of the symbol $\Lambda$ would then be a functional
$\hat{\Lambda}$.  However, with such a notion of model, some non
provable propositions are valid in all models, e.g. the extensionality
scheme.  Indeed, if $t$ and $u$ are two terms containing the free
variable $x$ such that $\fa x~(t = u)$ is valid, then $\llbracket t
\rrbracket_{a/x}$ and $\llbracket u \rrbracket_{a/x}$ are always
equal. Hence the terms $t$ and $u$ denote the same function, and thus
the terms $\Lambda x~t$ and $\Lambda x~u$ have the same denotation.
As we want to have the choice to take this extensionality scheme or
not, it should be valid in some models, but not all.  In other words,
extensionality should not be wired in the notion of model.  Thus we
need to consider a more general notion of model, leaving room to
interpret a term containing a free variable as an ``intensional
function''.

When, in the term $\Lambda x~t$, the body $t$ contains other free
variables than $x$, we interpret it as an intensional function of
several arguments. Thus, to be able to define the denotation of the
term $\Lambda x~t$, we must be able to apply the functional
$\hat{\Lambda}$ to intensional functions of various arities. Hence, we
shall define the denotation of $\Lambda$ as a family of functionals
$\hat{\Lambda}_{0}$, $\hat{\Lambda}_{1}$, $\hat{\Lambda}_{2}$, \ldots
each of these taking as arguments functions of some arity.

Once the notion of model is defined, we enter the proof of the
completeness theorem. The proof proceeds by giving a
provability preserving translation from binding logic to a formulation
of predicate logic, called {\em Deduction modulo}. This translation is
based on de Bruijn indices and explicit substitutions. Then, we relate
the models of a theory in binding logic with those of its translation
in Deduction modulo. It should be noticed that de Bruijn indices and
explicit substitutions are used in the proof of the completeness
theorem, but not in the definition of the notion of model
itself. Hence we can imagine other completeness proofs using other
methods such as a translation to higher-order logic, based on
higher-order abstract syntax or a direct construction of a model {\em
\`a la} Henkin. A potential difficulty in such a direct proof is to
construct the elements of the model, that have to be intensional, and
thus must differ from the functions used in Henkin's construction
\cite{Henkin,Andrews}.

\section{Binding logic}

\begin{definition}[Language]
A {\em language} is a set of function symbols and predicate symbols,
to each of which is associated a finite sequence of natural numbers
called its {\em binding arity} and denoted by $\langle k_{1}, \ldots , k_{n}
\rangle$ \cite{Pagano,Paganothese,FiorePlotkinTuri}.
\end{definition}

With respect of the standard vocabulary of predicate logic, a
function symbol or a predicate symbol of arity $n$ has binding arity 
$\barity{\underbrace{0, \ldots, 0}_n}$.

\begin{definition}[Terms]
Let ${\cal L}$ be a language and ${\cal V}$ be an infinite set of
variables. 
Terms are inductively defined as follows:
\begin{itemize}
\item variables are terms,
\item if $t_{1}, \ldots , t_{n}$ are terms, $f$ is a function symbol of
binding arity $\langle k_{1}, \ldots , k_{n} \rangle$, $x^{1}_{1}$, \ldots ,
$x^{1}_{k_{1}}$, \ldots , $x^{n}_{1}$, \ldots , $x^{n}_{k_{n}}$
are distinct variables, then:
$$f(x^{1}_{1} \cdots x^{1}_{k_{1}}~t_{1}, \ldots ,
x^{n}_{1} \cdots x^{n}_{k_{n}}~t_{n})$$
is a term.
\end{itemize}
\end{definition}

In the term 
$f(x^{1}_{1} \cdots x^{1}_{k_{1}}~t_{1}, \ldots ,
x^{n}_{1} \cdots x^{n}_{k_{n}}~t_{n})$
the variables $x^{i}_{1} \cdots x^{i}_{k_{i}}$ 
are simultaneously bound in the terms $t_{i}$.

A classical example is the binder of the lambda calculus; with
our notations, the lambda term $\lambda x~t$ is denoted $\lambda(x~ t)$ 
where the function symbol $\lambda$ is of binding arity $\barity{1}$.

\begin{definition}[Propositions]
Propositions are inductively defined as follows:
\begin{itemize}
\item if $t_{1}, \ldots , t_{n}$ are terms, $P$ is a predicate symbol of
binding arity $\langle k_{1}, \ldots , k_{n} \rangle$, $x^{1}_{1}$, \ldots ,
$x^{1}_{k_{1}}$, \ldots , $x^{n}_{1}$, \ldots , $x^{n}_{k_{n}}$ are
distinct variables then:
$$P(x^{1}_{1} \cdots x^{1}_{k_{1}}~t_{1}, \ldots ,
    x^{n}_{1} \cdots x^{n}_{k_{n}}~t_{n})$$
is a proposition,
\item if $A$ and $B$ are propositions, then $A \Rightarrow B$, $A
\wedge B$, $A \vee B$, $\bot$, $\fa x~A$ and $\ex x~A$ are
propositions.
\end{itemize}
\end{definition}

\VL{Free and bound variables are defined as usual.

\begin{definition}[Grafting]
Let $t$ be a term (resp. a proposition) and 
$\theta = t_{1}/x_{1}, \ldots , t_{n}/x_{n}$ 
be a finite function mapping variables
to terms. The term (resp. the proposition) $\langle
\theta\rangle t$ is defined as follows:
\begin{itemize}
\item $\langle \theta\rangle  x_{i} = t_{i}$,
\item $\langle \theta\rangle  x = x$ if $x$ is not in the domain of
$\theta$, 
\item $\langle \theta\rangle
f(x^{1}_{1} \cdots x^{1}_{k_{1}}~u_{1}, \ldots ,
  x^{n}_{1} \cdots x^{n}_{k_{n}}~u_{n})$

\hfill $= 
f(x^{1}_{1} \cdots x^{1}_{k_{1}}~\langle 
\theta_{|V \setminus \{x_{1}, ..., x_{n}\}}
\rangle u_{1}, \ldots ,
  x^{n}_{1} \cdots x^{n}_{k_{n}}~\langle 
\theta_{|V \setminus \{x_{1}, ..., x_{n}\}}
\rangle u_{n})$,

where $f$ is a function symbol or a predicate symbol,
\item $\langle \theta\rangle  (A \Rightarrow B) = \langle
\theta\rangle  A \Rightarrow \langle \theta\rangle  B $,
 $\langle \theta\rangle  (A \wedge B) = \langle \theta\rangle  A \wedge \langle
\theta\rangle  B $,\\
 $\langle \theta\rangle  (A \vee B) = \langle
\theta\rangle  A \vee \langle \theta\rangle  B $,
 $\langle
\theta\rangle  \bot = \bot$,\\
 $\langle \theta\rangle  \fa x~A = \fa
x~\langle \theta\rangle  A$,
 $\langle \theta\rangle  \ex x~A = \ex
x~\langle \theta\rangle  A$.
\end{itemize}
\end{definition}

Grafting allows captures. We introduce next substitution,
that avoids them. 

\begin{definition}[$\alpha$-equivalence]
The $\alpha$-equivalence relation is inductively defined as follows:
\begin{itemize}
\item $x \sim x$,
\item if $f$ is a function symbol or a predicate symbol, then:
$$f(x^{1}_{1} \cdots x^{1}_{k_{1}}~t_{1}, \ldots ,
   x^{n}_{1} \cdots x^{n}_{k_{n}}~t_{n}) \sim
f(x'^{1}_{1} \cdots x'^{1}_{k_{1}}~t'_{1}, \ldots ,
  x'^{n}_{1} \cdots x'^{n}_{k_{n}}~t'_{n})$$
if for all $i$
$\langle y_{1}/x^{i}_{1}, \ldots , y_{k_{i}}/x^{i}_{k_{i}} \rangle t_{i}
\sim
\langle y_{1}/x'^{i}_{1}, \ldots , y_{k_{i}}/x'^{i}_{k_{i}} \rangle
t'_{i}$
for a sequence of distinct variables $y_{1}, \ldots , y_{k_{i}}$ not appearing
(free or bound) in $t_{i}$ and $t'_{i}$,

\item
$A \Rightarrow B \sim A' \Rightarrow B'$,
$A \wedge B \sim A' \wedge B'$ and
$A \vee B \sim A' \vee B'$,
if $A \sim A'$ and $B \sim B'$,

\item
$\bot \sim \bot$,

\item $\fa x A \sim \fa x' A'$
and $\ex x A \sim \ex x' A'$ if
$\langle y/x\rangle A \sim \langle y/x'\rangle A'$ for a
variable $y$ not appearing
(free or bound) in $A$ and $A'$.
\end{itemize}
\end{definition}

It is easy to check that $\alpha$-equivalence is an equivalence
relation. From now on, terms will be considered up to
$\alpha$-equivalence.

\begin{definition}[Substitution]
\label{subst}
Let $t$ be a term (resp. a proposition) and 
$\theta = t_{1}/x_{1}, \ldots , t_{n}/x_{n}$ 
a finite function
mapping variables to terms. The
term (resp. the proposition) $\theta t$ is defined as follows:
\begin{itemize}
\item $\theta x_{i} = t_{i}$,
\item $\theta x = x$ if $x$ is not in the domain of $\theta$,
\item if $f$ is a function symbol or a predicate symbol,\\
$\theta
f(x^{1}_{1} \cdots x^{1}_{k_{1}}~u_{1}, \ldots ,  
  x^{n}_{1} \cdots x^{n}_{k_{n}}~u_{n}) =$

\hfill $f(y^{1}_{1} \cdots y^{1}_{k_{1}}~\theta \langle y^{1}_{1}/x^{1}_{1},
\ldots, y^{1}_{k_{1}}/x^{1}_{k_{1}} \rangle u_{1}, \ldots , y^{n}_{1} \cdots
y^{n}_{k_{n}}~\theta \langle y^{n}_{1}/x^{n}_{1}, \ldots ,
y^{n}_{k_{n}}/x^{n}_{k_{n}} \rangle u_{n})$ where $y^{1}_{1}$, \ldots ,
$y^{1}_{k_{1}}$, \ldots , $y^{n}_{1}$, \ldots , $y^{n}_{k_{n}}$ are
distinct variables appearing neither free nor bound neither in $f(x^{1}_{1}
\cdots x^{1}_{k_{1}}~u_{1}, \ldots , x^{n}_{1} \cdots x^{n}_{k_{n}}~
u_{n})$ nor in $\theta$.

\item $\theta (A \Rightarrow B) = \theta A \Rightarrow \theta B$,
$\theta (A \wedge B) = \theta A \wedge \theta B$,
$\theta (A \vee B) = \theta A \vee \theta B$,
$\theta \bot = \bot$,\\
$\theta \fa x~A = \fa y~\theta \langle y/x \rangle A$
where
$y$ is a variable appearing neither free nor bound neither in the
proposition nor in $\theta$,\\
$\theta \ex x~A = \ex y~\theta \langle y/x \rangle A$
where
$y$ is a variable appearing neither free nor bound neither in the
proposition nor in $\theta$.
\end{itemize}
\end{definition}

Notice that if $\theta = t_{1}/x_{1}, \ldots , t_{n}/x_{n}$ is a finite
function mapping variables to terms and $t$ a term or a proposition,
we write $\langle \theta \rangle t$ or $\langle t_{1}/x_{1}, \ldots ,
t_{n}/x_{n} \rangle t$ for the grafting of $\theta$ in $t$ and $\theta
t$ or $(t_{1}/x_{1}, \ldots , t_{n}/x_{n})t$ for the substitution of
$\theta$ in $t$.
}

\VC{
Free and bound variables, grafting (textual
replacement of a variable by a term, allowing 
captures), $\alpha$-conversion and substitution are defined as usual. 
If $\theta = t_{1}/x_{1}, \ldots , t_{n}/x_{n}$ is a finite
function mapping variables to terms and $t$ a term or a proposition,
we write $\langle \theta \rangle t$ or $\langle t_{1}/x_{1}, \ldots ,
t_{n}/x_{n} \rangle t$ for the grafting of $\theta$ in $t$ and $\theta
t$ or $(t_{1}/x_{1}, \ldots , t_{n}/x_{n})t$ for the substitution of
$\theta$ in $t$.
}

\begin{definition}[Sequent calculus] 
A sequent is a pair of multisets of propositions.  The sequent
calculus rules of {\em binding logic} are given figure
\ref{Seq}. There rules are the same than those of predicate logic,
except that, in the quantifier rules, the substitution avoids captures
not only with respect to quantified variables, but also with respect
to variables bound by the function and predicate symbols.
\end{definition}

{\begin{figure}
\hspace*{1.5cm}$
\begin{array}{ccc}

~~~~~~~~~~~~~~ & ~~~~~~~~~~~~~~~~~~~~~~~~~~~~ & \\

\irule{}
      {A \vdash A}
      {\mbox{axiom}}

& &
\irule{\Gamma, A \vdash \Delta ~~~ \Gamma \vdash A, \Delta}
      {\Gamma \vdash \Delta}
      {\mbox{cut}}
\\

\irule{\Gamma, A, A \vdash \Delta}
      {\Gamma, A \vdash \Delta}
      {\mbox{contr-left}}
& &
\irule{\Gamma \vdash A ,A ,\Delta}
      {\Gamma \vdash A,\Delta}
      {\mbox{contr-right}}
\\

\irule{\Gamma \vdash \Delta}
      {\Gamma, A \vdash \Delta} 
      {\mbox{weak-left}}
& &
\irule{\Gamma \vdash\Delta}
      {\Gamma \vdash A,\Delta}
      {\mbox{weak-right}}
\\

\irule{\Gamma \vdash A, \Delta ~~~ \Gamma, B \vdash \Delta}
      {\Gamma, A \Rightarrow B \vdash  \Delta}
      {\mbox{$\Rightarrow$-left}}
& &
\irule{\Gamma, A \vdash B, \Delta}
      {\Gamma \vdash  A \Rightarrow B, \Delta}
      {\mbox{$\Rightarrow$-right}}
\\

\irule{\Gamma, A, B \vdash \Delta}
      {\Gamma, A \wedge B \vdash  \Delta}
      {\mbox{$\wedge$-left}}
& &
\irule{\Gamma \vdash A, \Delta~~~\Gamma \vdash B, \Delta}
      {\Gamma \vdash A \wedge B, \Delta}
      {\mbox{$\wedge$-right}}
\\

\irule{\Gamma, A \vdash \Delta ~~~ \Gamma, B \vdash \Delta}
      {\Gamma, A \vee B \vdash  \Delta}
      {\mbox{$\vee$-left}}
&&
\irule{\Gamma \vdash A, B, \Delta}
      {\Gamma \vdash A \vee B, \Delta}
      {\mbox{$\vee$-right}}
\\

\irule{}
      {\Gamma, \bot \vdash \Delta}
      {\mbox{$\bot$-left}}
& &
\\

\irule{\Gamma, (t/x)A \vdash \Delta}
      {\Gamma, \fa x~A  \vdash \Delta}
      {\mbox{$\fa$-left}}
& &
\irule{\Gamma \vdash A, \Delta}
      {\Gamma \vdash \fa x~A, \Delta}
      {\mbox{$\fa$-right if $x \not\in FV(\Gamma \Delta)$}}
\\

\irule{\Gamma, A \vdash \Delta}
      {\Gamma, \ex x~A \vdash \Delta}
      {\mbox{$\ex$-left if $x \not\in FV(\Gamma \Delta)$}}
& &
\irule{\Gamma \vdash (t/x)A, \Delta}
      {\Gamma \vdash \ex x~A, \Delta}
      {\mbox{$\ex$-right}}
\end{array}
$
\caption{Sequent calculus of Binding Logic}
\label{Seq}
\end{figure}}

\begin{example}
The equational theory of $\lambda$-calculus can be defined in binding
logic as follows. The language contains a predicate $=$ of arity
$\langle 0,0 \rangle$, a function symbol $\alpha$ of arity $\langle
0,0 \rangle$ and a function symbol $\lambda$ of arity $\langle 1
\rangle$. 
We take the axioms of equality: reflexivity, symmetry, transitivity, 
compatibility with $\alpha$ and $\lambda$ \cite{Hindley}: 
$$\fa y~\fa z~(y = z \Rightarrow \lambda x~y = \lambda x~z)$$
and the axiom scheme $\beta$
$$\fa x~(\alpha(\lambda x~t,x) = t)$$ Notice here that we take a
different axiom for each term $t$, but that the argument of the
function is simply expressed by a variable.  

We may then add various extensionality axioms: the extensionality
scheme
$$(\fa x~(t = u)) \Rightarrow \lambda x~t = \lambda x~u$$ 
or the axiom $\eta$
$$\fa x~\fa y(\alpha(\lambda x~y,x) = y)$$
or the strong extensionality axiom 
$$(\fa x~(\alpha(f,x) = \alpha(g,x))) \Rightarrow f = g$$
yielding different theories.
\end{example}

\begin{example}
\label{delta}
The equational theory of the disjoint sum can be defined as follows. 
The language contains a predicate symbol $=$ of arity
$\langle 0, 0 \rangle$, function symbols $i$ and $j$ of arity $\langle
0 \rangle$ and $\delta$ of arity $\langle 0, 1, 1 \rangle$.
The theory is formed with the axioms of equality and the schemes
$$\delta(i(x),x~u,y~v) = u$$
$$\delta(j(y),x~u,y~v) = v$$
\end{example}

\begin{example}
An elementary theory of derivatives can be defined as follows.  Our
goal here is to stick to Leibniz notations, i.e. to differentiate
algebraic expressions with respect to some variable, not functions.
The language contains a predicate symbol $=$ of arity $\langle 0,0
\rangle$, constants $0$ and $1$ of arity $\langle \rangle$, function
symbols $+$ and $\times$ of arity $\langle 0,0 \rangle$ and a function
symbol $D$ of arity $\langle 1, 0 \rangle$.

The informal meaning is that $D(x~t,u)$ is the derivative of
the expression $t$ with respect to $x$ taken at $u$.  The term 
$D(x~t,u)$ is also written 
$\frac{\partial t}{\partial x}(u)$.

The axioms are 
$$\fa z~(\frac{\partial x}{\partial x}(x) = 1)$$
$$\fa x~(\frac{\partial y}{\partial x}(x) = 0)$$
and the schemes 
$$\fa x~(\frac{\partial (t+u)}{\partial x}(x) = 
\frac{\partial t}{\partial x}(x)
+ 
\frac{\partial u}{\partial x}(x))$$

$$\fa x~(\frac{\partial (t \times u)}{\partial x}(x) = 
t \times \frac{\partial u}{\partial x}(x)
+ 
\frac{\partial t}{\partial x}(x) \times u)$$
\end{example}

\section{Binding models}

As already discussed in the introduction, the denotation of a term
containing a free variable $x$ in a model ${\cal M}$ of domain $M$
must be an ``intensional function'' from $M$ to $M$. An intensional
function can be any mathematical object, provided it can be applied to
an element of $M$ and composed with other intensional functions.
Thus, besides $M$, we shall introduce in the definition of the model,
another set $M_{1}$ whose elements are called ``intensional
functions'' and functions from $M_{1} \times M$ to $M$, and from
$M_{1} \times M_{1}$ to $M_{1}$ for application and composition.

When a term $t$ contains more than one free variable, it must be
interpreted as an intensional function of several arguments. Hence, we
introduce also in the definition of the model sets $M_{2}$, $M_{3}$,
\ldots The elements of $M_n$ for $n \geq 1$ are called ``intensional
functions of 
$n$ arguments''. The set $M$ is then written $M_{0}$.  Among the
elements of $M_n$, if $n \geq 1$, we need the $n$ projections ${\bf
1}_{n}, \ldots, {\bf n}_{n}$. We also need functions to compose the
intensional functions and to apply them to elements of $M$. The
function $\Box_{0,n}$ applies an element of $M_n$ to a $n$-uple of
elements of $M_0$. The function $\Box_{p,n}$ composes an element of
$M_n$ with $n$ elements of $M_p$ and the result is an element of
$M_p$.

This leads to the following definition.

\begin{definition}[Intensional functional structure]
A {\em intensional functional structure} is a family of sets $M_{0},
M_{1}, M_{2}, \ldots$ 
together with
\begin{itemize}
\item 
for each natural number $n$, elements ${\bf 1}_{n}, \ldots, {\bf n}_{n}$
of $M_{n}$,
\item 
for each $n$ and $p$, a function $\Box_{p,n}$ from $M_{n} \times
M_{p}^{n}$ to $M_{p}$
\end{itemize}
verifying the properties 
$${\bf i}_{n} \Box_{p,n} \langle a_{1}, \ldots , a_{n} \rangle = a_{i}$$
$$a \Box_{n,n} \langle {\bf 1}_{n}, \ldots , {\bf n}_{n} \rangle =  a$$
$(a \Box_{p,n} \langle b_{1}, \ldots , b_{n} \rangle)
\Box_{q,p} \langle c_{1}, \ldots , c_{p} \rangle =$

\hfill 
$a \Box_{q,n}
\langle b_{1} \Box_{q,p} \langle c_{1}, \ldots , c_{p} \rangle
, \ldots ,  b_{n} \Box_{q,p} \langle c_{1}, \ldots , c_{p} \rangle
\rangle$
\end{definition}

\begin{remark}
Intensional function structures are multicategories with projections
(see \cite{Lambek69,Lambek89} and also \cite{Barber}) with a single
object $A$, the set $M_{n}$ being the set of multiarrows from $A^{n}$
to $A$.  If we wanted to define a model for many-sorted binding logic,
then we would need many-sorted intensional functional structures that
would exactly be multicategories with projections.
\end{remark}

\begin{example}
\label{example1}
Consider any set $A$.  Let $M_{n}$ be the functions from $A^{n}$ to
$A$,
\begin{itemize}
\item let ${\bf i}_{n}$ be the function mapping $\langle a_{1}, \ldots, a_{n}
\rangle$ to $a_{i}$, 
\item and $\Box_{p,n}$ be the function mapping $f$ (from $A^{n}$ to
$A$) and $g_{1}, \ldots, g_{n}$ (from $A^{p}$ to $A$) to the function 
mapping $\langle a_{1}, \ldots, a_{p} \rangle$ to 
$f(g_{1}(a_{1}, \ldots, a_{p}), \ldots, g_{n}(a_{1}, \ldots, a_{p}))$.
\end{itemize}
This structure is an intensional functional structure, as the properties are obviously
verified.
\end{example}

\begin{definition}[Binding model]
 
A model of a language ${\cal L}$ in binding logic is given by 
\begin{enumerate}
\item 
an intensional functional structure, 
\item for each function symbol of binding arity
$\langle k_{1}, \ldots, k_{n} \rangle $ a sequence of functions
$\hat{f}_{0}$, \ldots , $\hat{f}_{p}, \ldots $ with $\hat{f}_{p}$ 
in $(\prod_{i = 1}^n M_{p+k_i}) \rightarrow M_{p}$ such that
$$\hat{f}_{p}(a_{1}, \ldots , a_{n}) \Box_{q,p} 
\langle b_{1}, \ldots, b_{p} \rangle =
\hat{f}_{q}(a_{1} \Box_{q+k_{1},p+k_{1}}
\Uparrow_{q,k_{1},p}(\langle b_{1}, ..., b_{p} \rangle), \ldots)$$
where 
$$\Uparrow_{q,k,p}(\langle b_{1}, \ldots, b_{p} \rangle) = \langle {\bf
1}_{q+k}, \ldots, {\bf k}_{q+k},
b_{1} \Box_{q+k,q} S, \ldots, b_{p} \Box_{q+k,q} S \rangle$$
$$S = \langle {\bf (1+k)}_{q+k}, \ldots, {\bf (q + k)}_{q+k} \rangle$$
\item for each predicate symbol a function $\hat{P}$ in 
$(\prod_{i = 1}^n M_{k_i}) \rightarrow \{0,1\}$.
\end{enumerate}
\end{definition}

Notice that no variable can be bound outside a predicate symbol by a 
function symbol or a another predicate symbol. Hence 
the denotation of a predicate symbol is a single function and not a
family of functions. 

\begin{example}
If we have a single symbol $\Lambda$ of binding arity $\langle 1
\rangle$, then the denotation of this symbol is a family of functions 
$\hat{\Lambda}_{i}$ from $M_{i+1}$ to $M_{i}$ verifying, among others,
the equations
$$
\begin{array}{rl}
\hat{\Lambda}_{p}(a) & = (\hat{\Lambda}_{p}(a) \Box_{p+1,p} \langle {\bf 2},
\ldots, {\bf p + 1} \rangle) \Box_{p,p+1} \langle {\bf 1}, {\bf 1}, {\bf 2}, \ldots,
{\bf p} \rangle \\
 & = \hat{\Lambda}_{p+1}(a \Box_{p+2,p+1} \langle {\bf 1}, {\bf 3}, {\bf 4}, \ldots, {\bf
p+2} \rangle) \Box_{p,p+1} \langle {\bf 1}, {\bf 1}, {\bf 2}, \ldots,
{\bf p} \rangle
\end{array}
$$

Calling $I_{p}$ the function from $M_{p}$ to $M_{p+1}$ mapping $a$ to
$a \Box \langle {\bf 1}, {\bf 3}, \ldots, {\bf p+1} \rangle$
and $D_{p}$ the function from $M_{p+1}$ to $M_{p}$ mapping $a$ to
$a \Box \langle {\bf 1}, {\bf 1}, {\bf 2}, \ldots, {\bf p} \rangle$

We have 
$$\hat{\Lambda}_{p}(a) = D_{p}(\hat{\Lambda}_{p+1}(I_{p+1} (a)))$$

\begin{diagram}[height=1em,width=2em]
M_{1}&\rTo^{I_{1}}&M_{2}&\rTo^{I_{2}}&M_{3}&\rTo^{I_{3}}&M_{4}&\rTo^{I_{4}}&M_{5}&\rTo^{I_{5}}&\cdots\\
\\
\dTo^{\hat{\Lambda}_{0}}&&\dTo^{\hat{\Lambda}_{1}}&&\dTo^{\hat{\Lambda}_{2}}&&\dTo^{\hat{\Lambda}_{3}}&&\dTo^{\hat{\Lambda}_{4}}\\
\\
M_{0}&\lTo_{D_{0}}&M_{1}&\lTo_{D_{1}}&M_{2}&\lTo_{D_{2}}&M_{3}&\lTo_{D_{3}}&M_{4}&\lTo_{D_{4}}&\cdots\\
\end{diagram}
Each element in the family $\hat{\Lambda_{0}}, \hat{\Lambda}_{1},
\ldots$ adds information to the denotation of $\Lambda$. This
extension is coherent in the sense that the function $\Lambda_{p}$ can be
deduced from $\Lambda_{p+1}$.
\end{example}

\begin{remark}
This definition is reminiscent of that of \cite{FiorePlotkinTuri}
where a type constructor $\delta$ for context extension is
introduced. Then, a function symbol of binding arity $\langle k_{1},
\ldots, k_{n} \rangle$ is interpreted as a map $\delta^{k_{1}} A
\times \cdots \times \delta^{k_{n}} A \rightarrow A$. However our goal
here is somewhat different from that of \cite{FiorePlotkinTuri}, 
as it is to relate our notion of model to provability.
\end{remark}

\begin{definition} [Denotation in a model in binding logic]
The denotation of a term (resp. a proposition) in a model in binding
logic is
defined as follows. Let $x_{1}, \ldots , x_{p}$ be variables and
$\phi$ be an assignment mapping each free variable of $t$ but $x_{1},
\ldots, x_{p}$ to an element of $M_{0}$.
\begin{itemize}
\item
$\llbracket x_{k} \rrbracket_{\phi}^{x_{1}, \ldots , x_{p}} = {\bf k}_{p}$,
\item
$\llbracket y \rrbracket^{x_{1}, \ldots , x_{p}}_{\phi} = \phi(y) \Box_{p, 0}
\langle \rangle$ if $y$ is not one of the $x_{1}, \ldots , x_{p}$,
\item
$\llbracket f(y_{1}^1 \cdots y^1_{k_{1}}~t_{1}, \ldots ,
y^n_{1} \cdots y^n_{k_{n}}~t_{n}) \rrbracket^{x_{1}, \ldots , x_{p}}_{\phi}$

\hfill 
$= \hat{f}_{p}(\llbracket t_{1} \rrbracket^{y^1_{k_{1}}, \ldots ,
y^1_{1}, x_{1}, \ldots,
x_{p}}_{\phi}, \ldots,
\llbracket t_{n} \rrbracket^{y^n_{k_{n}}, \ldots , y^n_{1}, x_{1}, \ldots , x_{p}}_{\phi})$,
\item
$\llbracket P(y^1_{1} \cdots y^1_{k_{1}}~t_{1}, \ldots ,
    y^n_{1} \cdots y^n_{k_{n}}~t_{n}) \rrbracket_{\phi}
= \hat{P}(\llbracket t_{1} \rrbracket^{y^1_{k_{1}}, \ldots ,  y^1_{1}}_{\phi}, 
\ldots,
\llbracket t_{n} \rrbracket^{y^n_{k_{n}}, \ldots , y^n_{1}}_{\phi})$,
\item
$\llbracket A \Rightarrow B\rrbracket_{\phi} = \cdots$,
\item
$\llbracket \fa x~A \rrbracket_{\phi} = 1$ if for all $a$ in $M_{0}$,
$\llbracket A \rrbracket_{\phi + \langle x, a \rangle} = 1$,
\item
$\llbracket\ex x~A\rrbracket_{\phi} = 1$ if for some $a$ in $M_{0}$,
$\llbracket A \rrbracket_{\phi + \langle x, a \rangle} = 1$.
\end{itemize}
\end{definition}

\section{Examples of models}

\begin{example}
As an example, we give a model where the axioms of equality are valid,
but not the  extensionality scheme. The existence of such a model
will allow us to deduce that the  extensionality scheme is not a
consequence of the axioms of equality. We consider function symbols
$f$ of arity $\langle 0 \rangle$ and $\Lambda$ of arity $\langle 1
\rangle$, a predicate symbol $=$ of arity $\langle 0, 0 \rangle$. 
We build a model where the axioms
$$\fa x~(x = x)$$
$$\fa x \fa y~(x = y \Rightarrow y = x)$$
$$\fa x \fa y \fa z~(x = y \Rightarrow (y = z \Rightarrow x = z))$$
$$\fa x \fa y~(x = y \Rightarrow f(x) = f(y))$$
$$\fa x \fa y~(x = y \Rightarrow \Lambda z~x = \Lambda z~y)$$
are valid, but not the instance of the  extensionality scheme
$$(\fa x~f(x) = x) \Rightarrow \Lambda x~f(x) = \Lambda x~x$$ 

The base set $M_{0}$ of this model is a pair $\{k_{0}, l_{0}\}$.
The set $M_{n}$ contains the projections ${\bf 1}_{n},
\ldots, {\bf n}_{n}$, objects $\overline{1}_{n}, \ldots,
\overline{n}_{n}$ and two objects $k_{n}$ and $l_{n}$. Informally, the
objects $\overline{1}_{n}, \ldots, \overline{n}_{n}$ can be seen as
extensionally equal to, but distinct from, the projections and the
objects $k_{n}$ and $l_{n}$ as the constant functions taking the
values $k_{0}$ and $l_{0}$.

The model ${\cal M}$ is defined by $M_{n} = \{k_{n}, l_{n}, {\bf
1}_{n}, \ldots, {\bf n}_{n}, \overline{1}_{n}, \ldots,
\overline{n}_{n}\}$. Let $a$ be in $M_{n}$ and $b_{1}, \ldots,
b_{n}$ be in $M_{p}$, the element $a \Box_{p,n} \langle
b_{1}, \ldots, b_{n} \rangle$ of $M_{p}$ is defined as follows.
\begin{itemize}
\item if $a = k_{n}$ then 
$a \Box_{p,n} \langle b_{1}, \ldots, b_{n} \rangle = k_{p}$, 

\item if $a = l_{n}$ then $a \Box_{p,n} \langle b_{1}, \ldots, b_{n}
\rangle = l_{p}$, 

\item if $a = {\bf i}_{n}$ then $a \Box_{p,n} \langle b_{1}, \ldots, b_{n}
\rangle = b_{i}$, 

\item if $a = \overline{i}_{n}$ then $a \Box_{p,n} \langle b_{1}, \ldots,
b_{n} \rangle = - b_{i}$, where the involution $-$ is defined by
$$- k_{n} = k_{n}~~~- l_{n} = l_{n}~~~- {\bf i}_{n} = \overline{i}_{n}~~~
- \overline{i}_{n} = {\bf i}_{n}$$
\end{itemize}

\VC{We check
that 
$(-a) \Box_{p,n} \langle b_{1}, \ldots, b_{n} \rangle = - (a \Box_{p,n} 
\langle b_{1}, \ldots, b_{n} \rangle)$
by case analysis on $a$,
and then that ${\cal M}$ is an intensional functional structure.}

\VL{
\begin{proposition}
$(-a) \Box_{p,n} \langle b_{1}, \ldots, b_{n} \rangle = - (a \Box_{p,n} 
\langle b_{1}, \ldots, b_{n} \rangle)$
\end{proposition}

\proof{
\begin{itemize}
\item if $a = k_{n}$, both sides are equal to $k_{p}$,
\item if $a = l_{n}$, both sides are equal to $l_{p}$,
\item if $a = {\bf i}_{n}$, both sides are equal to $- b_{i}$,
\item if $a = \overline{i}_{n}$, both sides are equal to $b_{i}$.
\end{itemize}

}

\begin{proposition}
The structure ${\cal M}$ is an intensional functional structure. 
\end{proposition}

\proof{
First property:
$${\bf i}_{n} \Box_{p,n} \langle a_{1}, \ldots , a_{n} \rangle = a_{i}$$
By definition of $\Box$.\\
Second property:
$$a \Box_{n,n} \langle {\bf 1}_{n}, \ldots , {\bf n}_{n} \rangle =  a$$
\begin{itemize}
\item if $a = k_{n}$ then $a \Box_{n,n} \langle {\bf 1}_{n}, \ldots ,
{\bf n}_{n} \rangle =  k_{n}$, 
\item if $a = l_{n}$ then $a \Box_{n,n} \langle {\bf 1}_{n}, \ldots ,
{\bf n}_{n} \rangle =  l_{n}$, 
\item if $a = {\bf i}_{n}$ then $a \Box_{n,n} \langle {\bf 1}_{n}, \ldots ,
{\bf n}_{n} \rangle =  {\bf i}_{n}$, 
\item if $a = \overline{i}_{n}$ then $a \Box_{n,n} \langle {\bf
1}_{n}, \ldots , {\bf n}_{n} \rangle =  \overline{i}_{n}$.
\end{itemize}
Third property:\\
\noindent 
$(a \Box_{p,n} \langle b_{1}, \ldots , b_{n} \rangle)
\Box_{q,p} \langle c_{1}, \ldots , c_{p} \rangle =$

\hfill 
$a \Box_{q,n}
\langle b_{1} \Box_{q,p} \langle c_{1}, \ldots , c_{p} \rangle
, \ldots ,  b_{n} \Box_{q,p} \langle c_{1}, \ldots , c_{p} \rangle
\rangle$
\begin{itemize}
\item If $a = k_{n}$, both sides are equal to $k_{q}$,
\item if $a = l_{n}$, both sides are equal to $l_{q}$, 
\item if $a = {\bf i}_{n}$, both sides are equal to $b_{i} \Box \langle
c_{1}, \ldots, c_{q}\rangle$,
\item if $a = \overline{i}_{n} = - {\bf i}_{n}$, both sides are equal
to $- (b_{i} \Box \langle c_{1}, \ldots, c_{p} \rangle)$.
\end{itemize}}
}

We define the functions $\hat{f}_{n}$ from $M_{n}$ to $M_{n}$ by
$\hat{f}_{n}(x) = - x$ and the 
functions $\hat{\Lambda}_{n}$ from $M_{n+1}$ to $M_{n}$ by 
$$
\begin{array}{ll}
\hat{\Lambda}_{n}(k_{n+1}) = k_{n} & 
\hat{\Lambda}_{n}(l_{n+1}) = l_{n}\\
\hat{\Lambda}_{n}({\bf 1}_{n+1}) = k_{n} & 
\hat{\Lambda}_{n}(\overline{1}_{n+1}) = l_{n}

\\
\hat{\Lambda}_{n}({\bf i}_{n+1}) = {\bf (i-1)}_{n} \mbox{if $i \geq
2$} ~~~ & 
\hat{\Lambda}_{n}(\overline{i}_{n+1}) =
\overline{(i-1)}_{n} \mbox{if $i \geq 2$}
\end{array}
$$
At last, the function $\hat{=}$ from $M_{0} \times M_{0}$ to
$\{0,1\}$ is equality.

\VC{We check, 
that the families $\hat{f}_{i}$ and $\hat{\Lambda}_{i}$ are coherent,
i.e. that 
$$\hat{f}_{p}(a) \Box_{q,p} \langle b_{1}, \ldots, b_{p} \rangle =
\hat{f}_{q}(a \Box_{q,p} \langle b_{1}, \ldots, b_{p} \rangle)$$
$$\hat{\Lambda}_{p}(a) \Box_{q,p} \langle b_{1}, \ldots, b_{p} \rangle =
\hat{\Lambda}_{q}(a \Box_{q+1,p+1} \langle {\bf 1}_{q+1}, 
b_{1} \Box S, \ldots, b_{p} \Box S \rangle)$$
where $S = \langle {\bf 2}_{q+1}, \ldots, {\bf (q+1)}_{q+1} \rangle$,
by case analysis on $a$. To do so, we need to prove, by case analysis
of $b$ that $\hat{\Lambda}_{q}(b \Box S) = b$. 
}

\VL{
\begin{proposition}
The families $\hat{f}_{i}$ and $\hat{\Lambda}_{i}$ are coherent.
\end{proposition}

\proof{
For the case of the family $\hat{f}_{i}$, we have to prove that 
$$\hat{f}_{p}(a) \Box_{q,p} \langle b_{1}, \ldots, b_{p} \rangle =
\hat{f}_{q}(a \Box_{q,p} \langle b_{1}, \ldots, b_{p} \rangle)$$
\begin{itemize}
\item If $a = k_{p}$, both sides are equal to $k_{q}$,
\item if $a = l_{p}$, both sides are equal to $l_{q}$, 
\item if $a = {\bf i}_{p}$, both sides are equal to $- b_{i}$, 
\item if $a = \overline{i}_{p} = - {\bf i}_{p}$, both sides are equal
to $b_{i}$.
\end{itemize}

For the case of the family $\hat{\Lambda}_{i}$, we have to prove that 
$$\hat{\Lambda}_{p}(a) \Box_{q,p} \langle b_{1}, \ldots, b_{p} \rangle =
\hat{\Lambda}_{q}(a \Box_{q+1,p+1} \langle {\bf 1}_{q+1}, 
b_{1} \Box S, \ldots, b_{p} \Box S \rangle)$$
where $S = \langle {\bf 2}_{q+1}, \ldots, {\bf (q+1)}_{q+1} \rangle$. 

We first prove that if $b$ is an element of $M_{q}$ then 
$\Lambda_{q}(b \Box S) = b$.
\begin{itemize}
\item if $b = k_{q}$, both sides are equal to $k_{q}$, 
\item if $b = l_{q}$, both sides are equal to $l_{q}$, 
\item if $b = {\bf i}_{q}$, both sides are equal to ${\bf i}_{q}$, 
\item if $b = \overline{i}_{q}$, both sides are equal to $\overline{i}_{q}$.
\end{itemize}

Then we prove that
$$\hat{\Lambda}_{p}(a) \Box_{q,p} \langle b_{1}, \ldots, b_{p} \rangle =
\hat{\Lambda}_{q}(a \Box_{q+1,p+1} \langle {\bf 1}_{q+1}, 
b_{1} \Box S, \ldots, b_{p} \Box S \rangle)$$

\begin{itemize}
\item if $a = k_{p+1}$, both sides are equal to $k_{q}$,
\item if $a = l_{p+1}$, both sides are equal to $l_{q}$,
\item if $a = {\bf 1}_{p+1}$, then $a \Box_{q+1,p+1} \langle {\bf
1}_{q+1}, b_{1} \Box S, \ldots, b_{p} \Box S \rangle = {\bf 1}_{q+1}$ and
thus both sides are equal to $k_{q}$,
\item if $a = \overline{1}_{p+1}$, then $a \Box_{q+1,p+1} \langle {\bf
1}_{q+1}, b_{1} \Box S, \ldots, b_{p} \Box S \rangle = \overline{1}_{q+1}$ and
thus both sides are equal to $l_{q}$,
\item if $a = {\bf i}_{p+1}$ ($i \geq 2$), then using the property
above, both sides are equal to $b_{i-1}$,
\item if $a = \overline{i}_{p+1}$ ($i \geq 2$), then 
then using the property
above, both sides are equal to $- b_{i-1}$.
\end{itemize}}
}

\VC{As equality is interpreted by equality, the axioms of equality
are valid in ${\cal M}$.  The proposition $\fa x~f(x) = x$ is also
valid, but the proposition $\Lambda x~f(x) =
\Lambda x~x$ is not. Indeed $\llbracket \Lambda x~f(x) \rrbracket =
l_{0}$ and $\llbracket \Lambda x~x \rrbracket = k_{0}$.}

\VL{
\begin{proposition}
The axioms of equality are valid in the model above.
\end{proposition}

\proof{This is obvious as equality is interpreted by equality. 
We give the example of the axiom
$$\fa x \fa y~(x = y \Rightarrow \Lambda z~x = \Lambda z~y)$$
For every $a$ of $M_{0}$, we have
$\llbracket \Lambda z~x \rrbracket_{x/a} = \hat{\Lambda}_{0} (a \Box
\langle \rangle) = \llbracket \Lambda z~y \rrbracket_{a/y}$.}
}

\VL{
\begin{proposition}
The proposition
$$(\fa x~f(x) = x) \Rightarrow \Lambda x~f(x) = \Lambda x~x$$
is not valid in the model above.
\end{proposition}

\proof{
First we check that the proposition $\fa x~f(x) = x$
is valid, i.e. that for every $a$ we have 
$\llbracket f(x) \rrbracket_{a/x} = \llbracket x \rrbracket_{a/x}$.
This is a consequence of the fact that 
$\llbracket f(x) \rrbracket_{a/x} =
\hat{f}_{0}(a) = a = \llbracket x \rrbracket_{a/x}$.
Then we check that the proposition 
$\Lambda x~f(x) = \Lambda x~x$
is not valid, i.e. that 
$\llbracket \Lambda x~f(x) \rrbracket \neq \llbracket \Lambda x~x
\rrbracket$.
We have 
$\llbracket \Lambda x~f(x) \rrbracket = 
\hat{\Lambda}_{0} (\llbracket f(x) \rrbracket^{x}) = 
\hat{\Lambda}_{0} (\hat{f}_{1} (\llbracket x \rrbracket^{x})) = 
\hat{\Lambda}_{0} (\hat{f}_{1} ({\bf 1}_{1})) = 
\hat{\Lambda}_{0} (\overline{1}_{1}) = l_{0}$
and
$\llbracket \Lambda x~x \rrbracket = 
\hat{\Lambda}_{0} (\llbracket x \rrbracket^{x}) = 
\hat{\Lambda}_{0} ({\bf 1}_{1}) = k_{0}$ 
Thus 
$\llbracket \Lambda x~f(x) \rrbracket \neq \llbracket \Lambda x~x
\rrbracket$.}

\begin{proposition}
There are instances of the  extensionality scheme 
$$(\fa x~t = u) \Rightarrow \Lambda x~t = \Lambda x~u$$
that are not consequences of the axioms of equality.
\end{proposition}
}
\end{example}

\begin{example}
To the theory of example \ref{delta}, we add a constant $a$ of arity
$\langle \rangle$ and we want to prove that the proposition
$$\delta(a,x~a,y~a) = a$$
is not provable. 

We construct the following model. We take $M_{0} = {\mathbb
N}$, and for $n \geq 1$, $M_{n} = {\mathbb N}^n \ra {\mathbb
N}$.  The functions 
${\bf n}_{p}$ and $\Box_{n,p}$ are defined as in example
\ref{example1}. Thus the model is an intensional functional structure.
Then we take\\
\noindent
$\hat{a}_{0} = 1$ and accordingly $\hat{a}_{n} = x_{1}, \ldots, x_{n} \mapsto 
1$,\\
\noindent
$\hat{i}_{0} = x \mapsto 2 * x$ and accordingly 
$\hat{i}_{n} = x_{1}, \ldots, x_{n}, x \mapsto 2 * x$,\\
\noindent
$\hat{j}_{0} = x \mapsto 2 * x + 1$ and accordingly 
$\hat{j}_{n} = x_{1}, \ldots, x_{n}, x \mapsto 2 * x + 1$,\\
\noindent
$\hat{\delta}_{0} = d, f, g \mapsto 
\mbox{if $d = 0$ or $d = 1$ then $0$
else if $d$ is even then $f(d/2)$
else $g((d-1)/2)$}$,\\
and accordingly\\
$\hat{\delta}_{n} = x_{1}, \ldots, x_{n}, d, f, g \mapsto 
\hat{\delta}_{0}(d(x_{1}, \ldots x_{n}), y \mapsto f(x_{1}, \ldots, x_{n},y), 
y \mapsto g(x_{1}, \ldots, x_{n},y))$.\\
At last equality denotes equality.

The coherence properties are verified. The axioms of the theory are valid.
We have $\llbracket \delta(a, x~a, y~a) \rrbracket = 0$ and 
$\llbracket a \rrbracket = 1$.
Thus the proposition 
$$\delta(a, x~a, y~a) = a$$ is not valid. Hence it is not provable.

\end{example}

\VL{
\section{Deduction modulo}

We now want to prove that a sequent $\Gamma \vdash \Delta$ is provable
in binding logic if and only if for every model of $\Gamma$, there is
a proposition in $\Delta$ that is valid.  To prove this soundness and
completeness theorem, we shall translate binding logic into predicate
logic. We will prove that this translation preserves provability. Then
we shall relate the models of a theory in binding logic with the
models of its translation in predicate logic.

This translation is easier to define if we choose a formulation of
predicate logic where some axioms are replaced by computation
rules. This is the purpose of Deduction modulo introduced in
\cite{DHK-ENAR-98}. In this formalism, the notions of language, term
and proposition are that of predicate logic. But, a theory is formed
with a set of axioms $\Gamma$ {\em and a congruence $\equiv$} defined
on propositions.  As usual, the set $\Gamma$ must be decidable. The
congruence $\equiv$ must also be decidable and, in this paper,
congruences are defined by confluent and normalizing rewrite systems
on terms.

Deduction rules are modified to take this congruence into account. For
instance, the right rule of conjunction is not stated as usual:
$$\irule{\Gamma \vdash A, \Delta~~~\Gamma \vdash B, \Delta}
        {\Gamma \vdash A \wedge B, \Delta}
        {}$$
As its conclusion need not be exactly $A \wedge B$ but may be
only congruent to this proposition,  it is stated:
$$\irule{\Gamma \vdash A, \Delta~~~\Gamma \vdash B, \Delta}
        {\Gamma \vdash C, \Delta}
        {\mbox{if $C \equiv A \wedge B$}}$$
All rules of sequent calculus modulo are defined in a similar way (see
figure~\ref{SeqMod}).

\begin{figure}
$$\irule{}
        {A \vdash B}
        {\mbox{axiom if $A \equiv B$}}$$
$$\irule{\Gamma, A \vdash \Delta ~~~ \Gamma \vdash B, \Delta}
        {\Gamma \vdash \Delta}
        {\mbox{cut if $A \equiv B$}}$$
$$\irule{\Gamma, B_1, B_2 \vdash \Delta}
        {\Gamma, A \vdash \Delta}
        {\mbox{contr-left if $A \equiv B_1 \equiv B_2$}}$$
$$\irule{\Gamma \vdash B_1,B_2,\Delta}
        {\Gamma \vdash A,\Delta}
        {\mbox{contr-right if $A \equiv B_1 \equiv B_2$}}$$
$$\irule{\Gamma \vdash \Delta}
        {\Gamma, A \vdash \Delta}
        {\mbox{weak-left}}$$
$$\irule{\Gamma \vdash\Delta}
        {\Gamma \vdash A,\Delta}
        {\mbox{weak-right}}$$
$$\irule{\Gamma \vdash A, \Delta ~~~ \Gamma, B \vdash \Delta}
        {\Gamma, C \vdash  \Delta}
        {\mbox{$\Rightarrow$-left if $C \equiv (A \Rightarrow B)$}}$$
$$\irule{ \Gamma, A \vdash B, \Delta}
        {\Gamma \vdash  C, \Delta}
        {\mbox{$\Rightarrow$-right if $C \equiv (A \Rightarrow B)$}}$$
$$\irule{\Gamma, A, B \vdash \Delta}
        {\Gamma, C \vdash  \Delta}
        {\mbox{$\wedge$-left if $C \equiv (A \wedge B)$}}$$
$$\irule{\Gamma \vdash A, \Delta~~~\Gamma \vdash B, \Delta}
        {\Gamma \vdash C, \Delta}
        {\mbox{$\wedge$-right if $C \equiv (A \wedge B)$}}$$
$$\irule{\Gamma, A \vdash \Delta ~~~ \Gamma, B \vdash \Delta}
        {\Gamma, C \vdash  \Delta}
        {\mbox{$\vee$-left if $C \equiv (A \vee B)$}}$$
$$\irule{\Gamma \vdash A, B, \Delta}
        {\Gamma \vdash C, \Delta}
        {\mbox{$\vee$-right if $C \equiv (A \vee B)$}}$$
$$\irule{}
        {\Gamma, A \vdash \Delta}
        {\mbox{$\bot$-left if $A \equiv \bot$}}$$
$$\irule{\Gamma, C \vdash \Delta}
        {\Gamma, B \vdash \Delta}
        {\mbox{$(x,A,t)$~$\fa$-left if $B \equiv (\fa x~A)$ and $C \equiv (t/x)A$}}$$
$$\irule{\Gamma \vdash A, \Delta}
        {\Gamma \vdash B, \Delta}
        {\mbox{$(x,A)$~$\fa$-right if $B \equiv (\fa x~A)$ and
         $x \not\in FV(\Gamma \Delta)$}}$$
$$\irule{\Gamma, A \vdash \Delta}
        {\Gamma, B \vdash \Delta}
        {\mbox{$(x,A)$~$\ex$-left if $B \equiv (\ex x~A)$ and
         $x \not\in FV(\Gamma \Delta)$}}$$
$$\irule{\Gamma \vdash C, \Delta}
        {\Gamma \vdash B, \Delta}
        {\mbox{$(x,A,t)$~$\ex$-right if $B \equiv (\ex x~A)$  and 
        $C \equiv (t/x) A$}}$$
\caption{Sequent calculus modulo}
\label{SeqMod}
\end{figure}

Here is a very simple example, where the congruence on propositions is
defined by a rewrite system on terms. In arithmetic, we can define a
congruence with the following rules:
$$
\begin{array}{r@{~\rightarrow~}l}
0 + y & y\\
S(x) + y & S(x+y)\\
0 \times y & 0\\
S(x) \times y & x \times y + y\\
\end{array}
$$

In the theory formed with the axiom $\forall x~x = x$ and this
congruence,
we can prove, in sequent calculus modulo, that the number $4$ is even:
$$\irule{\irule{\irule{}
                      {4 = 4 \vdash 2 \times 2 = 4}
                      {\mbox{axiom}}
               }
               {\fa x~x = x \vdash 2 \times 2 = 4}
               {\mbox{$(x,x = x,4)$ $\forall$-left}}
        }
        {\fa x~x = x \vdash  \exists x~2 \times x = 4}
        {\mbox{$(x,2 \times x = 4,2)$ $\exists$-right}}$$
The sequent $4 = 4 \vdash 2 \times 2 = 4$, that requires a tedious proof
in usual formulations of arithmetic, can simply be proved with the axiom
rule here, as $(4 = 4) \equiv (2 \times 2 = 4)$.

\begin{proposition}
\label{memetaille}
If the sequent $A_{1}, \ldots , A_{n} \vdash B_{1}, \ldots , B_{p}$ has a
proof and 
$A_{1} \equiv A'_{1}$, \ldots, $A_{n} \equiv  A'_{n}$, $B_{1} \equiv B'_{1}$,
\ldots, $B_{p} \equiv B'_{p}$  then the sequent 
$A'_{1}, \ldots , A'_{n} \vdash B'_{1}, \ldots , B'_{p}$ has a
proof of the same length.
\end{proposition}

\proof{By induction over proof structure.}

\begin{definition}[Model of a theory modulo]
A model of a congruence is a model such that two congruent terms or
propositions have the same denotation.
A model of a theory modulo is a model of its axioms and of its congruence.
\end{definition}

G\"{o}del's completeness theorem generalizes to Deduction modulo. 

\begin{proposition}
[Soundness and completeness for Deduction modulo] 
(\cite{Dowek-habilitation})
A sequent $\Gamma \vdash \Delta$
is provable modulo a congruence $\equiv$ if and only if for every model of
$\langle \Gamma, \equiv \rangle$ there is a proposition of $\Delta$
that is valid. 
\end{proposition}
}

\section{Translation of binding logic into Deduction modulo}

\begin{definition}[Translation of a language]
Let ${\cal L}$ be a language in binding logic. We consider a
language ${\cal L}'$ in many-sorted predicate logic,
with sorts of the
form $n$ where $n$ is a natural number and $\langle n,p \rangle$ where
$n$ and $p$ are natural numbers. 
The sorts $n$ are those of terms appearing in a context where $n$
variables are bound. The sort $\langle n, p \rangle$ are those of
explicit substitutions mapping $p$ variables to $p$ terms, these terms
being of sort $n$.

To every function symbol $f$ of binding arity
$\langle k_{1}, \ldots , k_{n} \rangle$
we associate an infinite number
of function symbols $f_{0}, \ldots , f_{p}, \ldots $ of rank
$\langle k_{1} + p, \ldots , k_{n} + p \rangle p$.
To every predicate symbol $P$ of binding arity $\langle k_{1}, \ldots , k_{n}
\rangle$
we associate a predicate symbol also written $P$  of rank
$\langle k_{1}, \ldots , k_{n} \rangle$.
Then we introduce more constants and function symbols:
\begin{itemize}
\item for all natural numbers $n$, $n$ constants 
$1_{n}$, \ldots , $n_{n}$ of sort $n$,

\item for all natural numbers $n$ and $p$, a function symbol
$[]_{n,p}$ of rank $\langle p, \langle n,p \rangle \rangle n$,

\item for all natural numbers $n$, a constant $id_{n}$ of sort 
$\langle n,n \rangle$,

\item for all natural numbers $n$ and $p$, a function symbol 
$._{n,p}$ (read {\em cons}) of rank $\langle n, \langle n,p \rangle
\rangle \langle n,p+1 \rangle$,

\item for all natural numbers $n$, a constant  $\uparrow_{n}$ of
sort $\langle n+1,n \rangle$,

\item for all natural numbers $n$, $p$ and $q$, a function symbol
$\circ_{n,p,q}$ of rank $\langle \langle p,n \rangle,$ $\langle q,p
\rangle \rangle \langle q,n \rangle$.
\end{itemize}
\end{definition}

Notice that these symbols are those of the calculus of explicit
substitutions \cite{ACCL,CHL-JACM} with de Bruijn indices \cite{dB}.  
Calculi of explicit substitutions have first been introduced to 
decompose the substitutions initiated by $\beta$-reduction in
$\lambda$-calculus. It has then be noticed that having explicit 
renaming/relocating operators permitted to enforce scope constraints
in $\lambda$-calculus and then in other languages with bound variables
\cite{DHKunification,HOLSIGMA,Pagano,Paganothese,Vaillant}. 
In this paper, we are not particularly interested in $\beta$-reduction
and explicit substitutions are only used to enforce scope constraints.

We give now the translation from ${\cal L}$ to ${\cal L}'$. This
translation uses an auxiliary list of variables, which is written $l$. The length
of this list $l$ is written $|l|$ and, as usual, 
$\epsilon$ is the empty list, $x.l$ is the list $l$
prefixed by the term $x$ and $l_{p}$ is the $p$-th element of the list 
$l$.

\begin{definition}[Pre-cooking]
{\em Pre-cooking} is a function  from ${\cal L}$ to ${\cal L}'$ defined
as follows:
\begin{itemize}
\item
$F(x,l) = n_{|l|}$ if $l_{n} = x$ and $l_{p} \neq x$ for $p<n$,
\item
$F(x,l) = x[\uparrow_{0} \circ \uparrow_{1} \cdots \circ
\uparrow_{|l|-1}]$ if $l_{p} \neq x$ for all $p$. The variable $x$ is
given the sort $0$ in the language ${\cal L'}$. This term is simply
written $x[\uparrow^{|l|}]$ if there is no ambiguity,
\item
$F(f(x^{1}_{1} \cdots x^{1}_{k_{1}}~t_{1}, \ldots ,
     x^{n}_{1} \cdots x^{n}_{k_{n}}~t_{n}),l)$

\hfill $= f_{|l|}(F(t_{1}, x^{1}_{k_{1}} \cdots x^{1}_{1}.l), \ldots ,
F(t_{n}, x^{n}_{k_{n}} \cdots x^{n}_{1}.l))$,

\item
$F(P(x^{1}_{1} \cdots x^{1}_{k_{1}}~t_{1}, \ldots,
     x^{n}_{1} \cdots x^{n}_{k_{n}}~t_{n}), \epsilon)$

\hfill $=
P(F(t_{1}, x^{1}_{k_{1}} \cdots x^{1}_{1}), \ldots,
F(t_{n}, x^{n}_{k_{n}} \cdots x^{n}_{1}))$,

\item
$F(A \Rightarrow B,\epsilon) = F(A,\epsilon) \Rightarrow F(B,\epsilon)$,\\
$F(A \wedge B,\epsilon) = F(A,\epsilon) \wedge F(B,\epsilon)$,\\
$F(A \vee B,\epsilon) = F(A,\epsilon) \vee F(B,\epsilon)$,\\
$F(\bot,\epsilon) = \bot$,\\
$F(\fa x~A, \epsilon) = \fa x~F(A,\epsilon)$,\\
$F(\ex x~A, \epsilon) = \ex x~F(A,\epsilon)$.
\end{itemize}
We write $t'$ for $F(t,\epsilon)$ and $A'$ for $F(A,\epsilon)$. 
\end{definition}

\begin{proposition}
If $t$ is a term of ${\cal L}$ then $t'$ is a term of sort $0$.
\end{proposition}

\VL{
\proof{
We prove, by induction on terms, the more general proposition:
if $t$ is a term of ${\cal L}$ then $F(t,l)$ is a term of sort $|l|$.

\begin{itemize}
\item
If $t$ is a variable of $l$, then $F(t,l) = n_{|l|}$ and thus it has
sort $|l|$. 

\item
If $t$ is a variable $x$ not belonging to $l$, then 
$F(t,l) = x[\uparrow^{|l|}]$. 
The term
$\uparrow^{|l|} = \uparrow_{0} \circ \uparrow_{1} \cdots \circ
\uparrow_{|l|-1}$ has sort $\langle |l|, 0\rangle$. The variable $x$
has sort $0$. So, $x[\uparrow^{|l|}]$ has sort $|l|$.
\item If $t = f(x^{1}_{1} \cdots x^{1}_{k_{1}}~t_{1}, \ldots , x^{n}_{1}
\cdots x^{n}_{k_{n}}~t_{n})$ then 
$F(t,l) = f_{|l|}(F(t_{1}, x^{1}_{k_{1}} 
\cdots x^{1}_{1}.l), \ldots, F(t_{n}, x^{n}_{k_{n}} \cdots x^{n}_{1}.l))$.
By induction hypothesis, $F(t_{i}, x^{i}_{k_{i}} \cdots x^{i}_{1}.l)$ have
sort $k_{i} + |l|$ and the application of the symbol $f_{|l|}$ 
yields a term of sort $|l|$.
\end{itemize}
}
}
\begin{example}
Let $f$ be a function symbol of binding arity $\langle 0, 0 \rangle$ and
$\Lambda$ a 
function symbol of binding arity $\langle 1 \rangle$ and $=$ a predicate
symbol of binding arity $\langle 0, 0 \rangle$. The proposition:
$$\fa x~\fa y~(f(x,y) = \Lambda z.f(x,z))$$
is translated as:
$$\fa x~\fa y~(f_{0}(x,y) = \Lambda_{0}(f_{1}(x[\uparrow],1_{1})))$$

The variables $x$ and $y$, that are bound by quantifiers, are kept as
named variables, the variable $z$, that is bound by a binding symbol
is replaced by a de Bruijn index, the first occurrence of $f$ is
translated as $f_{0}$ and the second as $f_{1}$ as it is under one
binder, the variable $x$ under the binder $\Lambda$ is protected by a
substitution $\uparrow$ to avoid captures. The terms $x[\uparrow]$,
$1_{1}$ and $f_{1}(x[\uparrow],1_{1})$ have sort $1$ while the term
$\Lambda_{0}(f_{1}(x[\uparrow],1_{1}))$ has sort $0$.
\end{example}

\begin{definition}[Translation of a theory]

Let $\Gamma$ be a theory in binding logic. The translation of
$\Gamma$ is a theory modulo having as axioms the translations of those
of $\Gamma$ and as congruence that defined by the following rewrite
system $\sigma$: 
$$\begin{array}{rcl}
n + 1 & ~\rightarrow~ & 1[\uparrow^{n}]\\
1[t.s] & ~\rightarrow~ & t\\
t[id] & ~\rightarrow~ &  t\\
(t[s])[s'] & ~\rightarrow~ &  t[s \circ s']\\
id \circ s & ~\rightarrow~ &  s\\
\uparrow \circ (t.s) & ~\rightarrow~ &  s\\
(s_1 \circ s_2) \circ s_3 & ~\rightarrow~ &  s_1 \circ (s_2 \circ s_3)\\
(t.s) \circ s' & ~\rightarrow~ &  t[s'].(s \circ s')\\
s \circ id & ~\rightarrow~ &  s\\
1. \uparrow & ~\rightarrow~ &  id\\
1[s].(\uparrow \circ ~ s) & ~\rightarrow~ & s\\
f_{p}(t_{1}, \ldots , t_{n})[s] & ~\rightarrow~ & f_{q}(t_{1}[
1. 1[\uparrow]. \cdots . 1[\uparrow^{k_1-1}].s \circ \uparrow^{k_{1}}], \\
 & & \rule{5mm}{0mm} \ldots,\\
 & & \rule{5mm}{0mm} t_{n}[1.1 [\uparrow]. \cdots .  1[\uparrow^{k_n-1}].s \circ \uparrow^{k_{n}}])
\end{array}$$
where, in the last rule, $f$ has binding arity $\langle k_{1}, \ldots,
k_{n} \rangle$ and $s$ has sort $\langle q, p \rangle$. 
\end{definition}

The system $\sigma$ is locally confluent as all critical pairs
converge.  To obtain termination, it suffices to note that $\sigma$
shares all the rules but the first and the last ones with the system
$\sigma$ of \cite{ACCL,CHL-JACM}, called $\sigma_{e}$ here, which is
strongly normalizing. Here we use a version of $\sigma_e$ with pairs,
which is also strongly normalizing \cite{CHL-JACM}.  For the first and
the last rule, we translate $n+1$ onto $1[\uparrow ^n] [id]$ and
$f_{p}(t_{1}, \ldots , t_{n})$ onto $\langle \cdots \langle
\lambda^{k_1} t_1, \lambda^{k_2} t_2 \rangle, \ldots, \lambda^{k_n}
t_n \rangle$. The right member of the last rule translates onto the
term $\langle \cdots \langle \lambda^{k_1} t_1[1.1[\uparrow]. \cdots
. 1[\uparrow^{k_{1} -1}] .s \circ \uparrow^{k_{1}}], \ldots \rangle,
\ldots, \lambda^{k_n} t_n [1. 1[\uparrow] \cdots
. 1[\uparrow^{k_{n}-1}] .s \circ \uparrow^{k_{n}}] \rangle$, which is
the reduct by rules of $\sigma_e$ of the translation of the left
member. Thus, every rewriting step of $\sigma$ is translated as a non
empty sequence of rewriting steps of $\sigma_e$. As $\sigma_e$ is
strongly normalizing, this is also the case for $\sigma$. 

If $t$ is a term or a proposition, we write $t \downarrow$ for its
normal form. 

\VL{

We now want to prove that the sequent $\Gamma \vdash \Delta$ is
provable in binding logic if and only if $\Gamma' \vdash \Delta'$ is
provable in Deduction modulo $\sigma$.
First, we show that substitution (with renaming) in ${\cal L}$ 
collapses to grafting (without renaming) in ${\cal L}'$. 

\begin{proposition}
\label{substterm}

Let $t$ and $u$ be terms of ${\cal L}$. Then:
$$((t/x)u)' \equiv \langle t'/x \rangle u'$$
\end{proposition}

\proof{To ease reading, we leave implicit the sorts of the symbols.
We prove by induction on the structure of $u$ the more general
statement: for all lists $l$, if the free variables of $t$ do not
appear in $l$, then:
$$F((t/x)u,l) = \langle t'/x \rangle F(u,l)$$
The result follows for $l = \epsilon$. 
The non trivial case is when $u$ is $x$, in this case we need:
$$F(t,l) = \langle t'/x \rangle F(x,l)$$
i.e.:
$$F(t,l) = t'[\uparrow^{|l|}] = F(t,\epsilon)[\uparrow^{|l|}]$$
To prove this, we show by induction on the structure of $t$, that
for all lists $l$ and $l'$ such that the free variables of $t$ do not appear
in $l$, we have:
$$F(t,l'~l) = F(t,l')[1. \cdots |l'|. \uparrow^{|l|+|l'|}]$$
 where $l'~l$ is the concatenation of the two lists.

\begin{itemize}
\item If $t$ is a variable of $l'$, then $t$ is the $k^{th}$ variable
of $l'$ for some $k$ and both terms reduce to $k$.

\item If $t$ is a variable $x$ that does not belong to $l'$ then 
it does not belong to $l'~l$ and both terms reduce to
$x[\uparrow^{|l|+|l'|}]$. 

\item If $t = f(x^{1}_{1} \cdots x^{1}_{k_{1}}~t_{1}, \ldots ,
x^{n}_{1} \cdots x^{n}_{k_{n}}~t_{n})$
then, expliciting only the first argument of $f$ for the sake of simplicity,
 we get: 
$$F(t,l'~l) =
f_{|l'|+|l|} (F(t_{1}, x^{1}_{k_{1}} \cdots x^{1}_{1}~l'~l), \ldots )$$
and by induction hypothesis:
$$F(t,l'~l) =
f_{|l'|+|l|} (F(t_{1}, x^{1}_{k_{1}} \cdots x^{1}_{1}~l')
[1. \cdots |l'| + k_{1}. \uparrow^{|l|+|l'|+k_{1}}]
, \ldots 
)$$
Then, we have:
$$F(t,l')[1. \cdots |l'|. \uparrow^{|l|+|l'|}] =
f_{|l'|} (F(t_{1}, x^{1}_{k_{1}} \cdots x^{1}_{1}~l'), \ldots )
[1. \cdots |l'|. \uparrow^{|l|+|l'|}]$$
and with the last rule of $\sigma$, we get:

\noindent 
$F(t,l')[1. \cdots |l'|. \uparrow^{|l|+|l'|}] =$

$\hfill f_{|l'|+|l|} (
F(t_{1}, x^{1}_{k_{1}} \cdots x^{1}_{1}~l')
[1. \cdots |l'| + k_{1}. \uparrow^{|l| + |l'| + k_{1}}]
, \ldots )$

\noindent
Thus:
$$F(t,l'~l) = F(t,l')[1. \cdots |l'|. \uparrow^{|l|+|l'|}]$$
\end{itemize}
} 

\begin{proposition}
\label{substprop}
Let $t$ and $A$ be a term and a proposition of ${\cal L}$. Then:
$$((t/x)A)' \equiv (t'/x)A'$$
\end{proposition}

\proof{By induction over the structure of $A$. 
If $A$ is atomic then we apply the proposition \ref{substterm}. Notice
that if $u$ is a term without bound variables, then $(t/x)u = \langle t/x
\rangle u$.
If $A$ has the form $B \Rightarrow C$, $B \wedge C$, $B \vee C$ or
$\bot$, we apply the induction hypothesis. 
If $A$ has the form $\fa y~B$, then let $z$ be a fresh variable, we
have:
$$(t/x)A = (t/x)(\fa y~B) = \fa z~((t/x) \langle z/y\rangle B)$$ 
Thus $((t/x)A)' = \fa z~((t/x) \langle z/y \rangle B)' 
\equiv \fa z~(t'/x)( \langle z/y \rangle B)'
= (t'/x) \fa z~(\langle z/y \rangle B)'$
$= (t'/x) \fa y~B' = (t'/x) A'$. 
If $A$ has the form $\ex y~B$, then the proof is similar.}

\medskip

The purpose of the following definition and proposition is to
characterize the image of pre-cooking.

\begin{definition}[$F$-term]
A {\em $F$-term} (resp. a {\em $F$-proposition}) is a $\sigma$-normal
term (resp. proposition) of ${\cal L}'$ containing 
only variables of sort $0$. 
\end{definition}

\begin{proposition}
\label{image}
If $t$ is a $F$-term of sort $0$ (resp. a $F$-proposition), then there
is a term (resp. a proposition) $u$ in ${\cal L}$  such that $t = u'$.
\end{proposition}

\proof{We first prove by induction on the structure of $t$ that if $t$
is a $F$-term of sort $n$ then there is a term $u$ and a sequence $l$
of $n$ variables such that $t = F(u,l)$.
The non trivial case is when $t = x[s]$. This term has sort
$n$ thus $s$ has sort $\langle n,0 \rangle$ and it is normal, thus
$s = \uparrow^{n}$. 

Then, we prove by induction on the structure of $A$ that if $A$ is a
$F$-proposition then there is a proposition $B$ 
such that $A = B'$.}

\begin{proposition} \label{key}
Let $\Gamma \vdash \Delta$ be a sequent containing only
$F$-propositions.  If the sequent $\Gamma \vdash \Delta$ has a proof,
then it also has a proof where all propositions are $F$-propositions
and all the witnesses $F$-terms of sort $0$.
\end{proposition}

\proof{First, as $\sigma$ is a term rewrite system, Deduction modulo
$\sigma$ has the cut elimination property \cite{DowekWerner} and thus
$\Gamma \vdash \Delta$ has a cut free proof.  Then we reason by
induction on the size of such a cut free proof.

If the last rule is an axiom then the result is obvious. Otherwise, we
apply the induction hypothesis to the sub-proofs. Consider for
instance the case of the right rule of the existential quantifier:
$$\irule{\irule{\pi}
               {\Gamma \vdash C, \Delta}
               {}
        }
        {\Gamma \vdash B, \Delta}
        {\mbox{$(x,A,t)$ $\ex$-right}}$$
With $B \equiv \ex x~A$ and $C \equiv (t/x)A$.

Call $\theta$ the substitution mapping each variable $y$ of $t$
of sort $n$ ($n \neq 0$) to the term $y'[\uparrow^{n}]$
where $y'$ is a fresh variable of sort $0$.
By induction on the structure of $\pi$, the proof 
$\theta \pi$ is a proof of $\Gamma \vdash \theta C, \Delta$
we have $\theta C \equiv \theta (t/x) A = (\theta t / x) A \equiv 
((\theta t /x) A )\downarrow$. 
Thus, by proposition \ref{memetaille},
there is a proof $\pi'$ of the same size as $\pi$ of the sequent 
$\Gamma \vdash ((\theta t/y)A) \downarrow, \Delta$.

The sequent $\Gamma \vdash ((\theta t/y)A) \downarrow, \Delta$ contains
only $F$-propositions. Applying the induction hypothesis to the proof
$\pi'$, we get that there exists a proof $\pi''$ of
$\Gamma' \vdash ((\theta t/y)A) \downarrow, \Delta$ where all
propositions are $F$-propositions and all the witnesses $F$-terms of
sort $0$.
We build the proof:
$$\irule{\irule{\pi''}
               {\Gamma \vdash ((\theta t/y)A) \downarrow, \Delta }
               {}
        }
        {\Gamma \vdash B, \Delta}
        {\mbox{$(x,A,\theta t)$ $\ex$-right}}$$}
} 

\VC{

We now want to prove that the sequent $\Gamma \vdash \Delta$ is
provable in binding logic if and only if $\Gamma' \vdash \Delta'$ is
provable in Deduction modulo $\sigma$.

The fact that the sequent $\Gamma' \vdash \Delta'$ is proved in
Deduction modulo $\sigma$ and not in usual predicate logic simplifies
the proof of this result.

\begin{definition}[$F$-term]
A {\em $F$-term} (resp. a {\em $F$-proposition}) is a $\sigma$-normal
term (resp. proposition) of ${\cal L}'$ containing 
only variables of sort $0$. 
\end{definition}

\begin{proposition}
\label{substterm}
\label{substprop}
\label{image}
\label{key}

\begin{enumerate}
\item
Let $t$ and $u$ be terms of ${\cal L}$. Then:
$$((t/x)u)' \equiv \langle t'/x \rangle u'$$

\item
Let $t$ and $A$ be a term and a proposition of ${\cal L}$. Then:
$$((t/x)A)' \equiv (t'/x)A'$$

\item
If $t$ is a $F$-term of sort $0$ (resp. a $F$-proposition), then there
is a term (resp. a proposition) $u$ in ${\cal L}$  such that $t = u'$.

\item
Let $\Gamma \vdash \Delta$ be a sequent containing only
$F$-propositions.  If the sequent $\Gamma \vdash \Delta$ has a proof,
then it also has a proof where all propositions are $F$-propositions
and all the witnesses $F$-terms of sort $0$.
\end{enumerate}
\end{proposition}

} 

\VC{We can now state the theorem:}
\VL{We can now state and prove the theorem:}

\begin{theorem} 
\label{main}
The sequent $\Gamma \vdash \Delta$ is provable in binding logic if and
only if $\Gamma' \vdash \Delta'$ is provable modulo $\sigma$.
\end{theorem}

\VL{\proof{
The direct sense is an easy induction on the structure of the proof in
binding logic. As an example, we give  the case of the
right rule of the existential quantifier.
The proof has the form:
$$\irule{\irule{\pi}
               {\Gamma \vdash (t/x)A, \Delta}
               {}
        }
        {\Gamma \vdash \ex x~A}
        {\mbox{$\ex$-right}}$$
By induction hypothesis, there is a proof $\pi'$ of
$\Gamma' \vdash ((t/x)A)', \Delta'$,
and, by proposition \ref{substprop}, we have
$(t'/x)A' \equiv ((t/x)A)'$.
We build the proof in Deduction modulo:
$$\irule{\irule{\pi'}
               {\Gamma' \vdash ((t/x)A)', \Delta'}
               {}
        }
        {\Gamma' \vdash \ex x~A', \Delta'}
        {\mbox{$(x,A',t')$ $\ex$-right}}$$
Conversely, by the proposition \ref{key}, we can build a proof in
Deduction modulo of $\Gamma' \vdash \Delta'$ where all the
propositions are $F$-propositions and all the witnesses
$F$-terms of sort $0$. By proposition \ref{image}, all the
propositions and witnesses in this proof are the pre-cooking of some
propositions or terms in binding logic. 
By induction on the structure of this proof we can build a
proof of $\Gamma \vdash \Delta$ in binding logic. As an example, we
give  the case of the right rule of the existential quantifier.
The proof has the form:
$$\irule{\irule{\pi}
               {\Gamma' \vdash C', \Delta'}
               {}
        }
        {\Gamma' \vdash B', \Delta'}
        {\mbox{$(x,t',A')$ $\ex$-right}}$$
where $B' \equiv \ex x~A'$ and $C' \equiv (t'/x)A'$.

By proposition \ref{substprop}, we have
$(t'/x)A' \equiv ((t/x)A)'$ and thus $C' \equiv ((t/x)A)'$.
As $F$-terms are $\sigma$-normal we have
$B' = \ex x~A'$ and $C' = ((t/x)A)'$.
As pre-cooking is injective, we get $B = \ex x~A$ and $C = (t/x)A$.

By induction hypothesis, there exists a proof $\pi'$ in binding logic
of $\Gamma \vdash (t/x)A, \Delta$. We build the proof in binding logic:
$$\irule{\irule{\pi'}
               {\Gamma \vdash (t/x)A, \Delta}
               {}
        }
        {\Gamma \vdash \ex x~A, \Delta}
        {\mbox{$\ex$-right}}$$
}
}
Thus, binding logic can be coded back into predicate logic
modulo. We could pursue this by coding back predicate logic modulo
into predicate logic (introducing the rules of $\sigma$ as
axioms). However, this is not necessary to define the notion of model,
because we can use directly the notion of model of Deduction modulo
and the related soundness and completeness result.

\begin{remark}
An axiom scheme, containing schematic variables for terms 
such as the scheme $\beta$
$$\fa x~(\alpha(\lambda x~t,x) = t)$$
is translated as an axiom scheme
$$\fa x~(\alpha(\lambda F(t,x.\epsilon),x) = F(t,\epsilon))$$
Using the fact that $F(t,x.\epsilon)[x.id] = F(t,\epsilon)$ we can replace this
scheme by the equivalent 
$$\fa x~(\alpha(\lambda F(t,x.\epsilon),x) = F(t,1)[x.\epsilon])$$
i.e. by 
$$\fa x~(\alpha(\lambda u,x) = u[x.\epsilon])$$
where $u$ is a schematic variable ranging over terms of the form
$F(t,x.\epsilon)$. 

However, we may also imagine to translate it as a single axiom
$$\fa Y \fa x~(\alpha(\lambda Y,x) = Y[x.id])$$
Notice that the variable $Y$ has sort $1$ here, while $x$ has sort
$0$.

This remark shows that our method for proving completeness could be
adapted for extensions of binding logic with quantification over
variables ranging in $M_{i}$ for $i \geq 1$. 
We leave this for future work.
\end{remark}

\section{Relating models}

\begin{proposition}
\label{a}
The sequent $\Gamma \vdash \Delta$ is provable in binding logic if 
and only if for every model of $\langle \Gamma', \equiv \rangle$, there is a
proposition in $\Delta'$ that is valid.
\end{proposition}

\VL{\proof{The sequent $\Gamma \vdash \Delta$ is provable in binding
logic if and only if the sequent $\Gamma' \vdash_{\equiv} \Delta'$ is
provable in Deduction modulo. By the soundness and completeness
theorem of Deduction modulo this is equivalent to the fact that for
every model of $\langle \Gamma', \equiv \rangle$, there is a
proposition in $\Delta'$ that is valid.}}

Now, we want to prove that for every model of $\Gamma$ there is a
proposition in $\Delta$ that is valid if and only if for every model
of $\langle \Gamma', \equiv \rangle$ there is a proposition in
$\Delta'$ that is valid.  This is the purpose of the two next
propositions that form a collapse lemma, showing that interpreting
${\cal L'}$ we can restrict to models where explicit substitutions are
interpreted like sequences of terms.

\begin{proposition}
\label{b}
If for every model of $\langle \Gamma',\equiv \rangle$, there is a
proposition in $\Delta'$ that is valid, then 
for every model of $\Gamma$, there is a
proposition in $\Delta$ that is valid.
\end{proposition}

\VL{
\proof{Let ${\cal M}$ be a model of $\Gamma$ (in binding logic). We
want to prove that some proposition of $\Delta$ is valid in ${\cal M}$. 

We build a model ${\cal N}$ of ${\cal L}'$ in Deduction modulo. 
We take $N_{n} = M_{n}$, $N_{n,p} = M_{n}^{p}$ (the Cartesian product
of $M_{n}$), 
\begin{itemize}
\item $\hat{p}_{n} = {\bf p}_{n}$, 
\item $a \hat{[]}_{n,p} \langle b_{1}, \ldots, b_{p} \rangle = a
\Box_{n,p} \langle b_{1}, \ldots, b_{p} \rangle$, 

\item $\hat{id}_{n} = \langle {\bf 1}_{n}, \ldots , {\bf n}_{n} \rangle$,

\item $\hat{.}_{n,p}(a, \langle b_{1}, \ldots , b_{p} \rangle) =
\langle a, b_{1}, \ldots , b_{p} \rangle$,
\item
$\hat{\uparrow}_{n} = \langle {\bf 2}_{n+1}, \ldots , {\bf
 (n+1)}_{n+1} \rangle$ 
 \item $\hat{\circ}_{n,p,q} (\langle a_{1}, \ldots , a_{p} \rangle,
\langle b_{1}, \ldots , b_{q} \rangle)
= \langle a_{1} \Box_{n,q} \langle b_{1}, \ldots , b_{q} \rangle,
\ldots,
a_{p} \Box_{n,q} \langle b_{1}, \ldots , b_{q} \rangle \rangle$
\end{itemize}
and we keep the same denotation for the function symbols and the
predicate symbols.

Then we verify that the model ${\cal N}$ is a model of the rewrite
rules of $\sigma$. This is a consequence of the definition of the
denotation of the symbols of ${\cal L'}$ and of the fact that ${\cal
M}$ is a model in binding logic (hence it verifies the three
properties of intensional functional structures and the coherence property for
function symbols). 

Then, a simple proof by induction over the structure of terms
(resp. propositions) shows that a the denotation of $t$ in ${\cal M}$
is the denotation of $t'$ in ${\cal N}$. 

Hence, the model ${\cal N}$ is is a model of $\Gamma'$. As it is a
model of $\Gamma'$ and $\equiv$, there is a proposition in $\Delta'$
that is valid in ${\cal N}$. Thus, there is a proposition of $\Delta$
that is valid in ${\cal M}$.}
}

\begin{proposition}
\label{c}

If for every model of $\Gamma$, there is a
proposition in $\Delta$ that is valid, then 
for every model of $\langle \Gamma', \equiv \rangle$
there is a proposition in $\Delta'$ that is valid.
\end{proposition}

\VL{\proof{Let ${\cal N}$ be a model of $\langle \Gamma', \equiv \rangle$
(in deduction 
modulo). We want to prove that some proposition of $\Delta'$ is valid
in ${\cal N}$.  

The model ${\cal N}$ is formed with a family of sets $N_{n}$,
a family of sets $N_{n,p}$ and functions 
$\hat{p}_{n}$, $\hat{[]}_{n,p}$, $\hat{id}$, $\hat{.}_{n,p}$,
$\hat{\uparrow}_{p}$, $\hat{\circ}_{n,p,q}$. 

We define a model ${\cal M}$ of ${\cal L}$ (in binding logic)
by taking $M_{n} = N_{n}$, 
\begin{itemize}
\item ${\bf p}_{n} = \hat{p}_{n}$,
\item $a {\Box}_{n,p} \langle b_{1}, \ldots, b_{p} \rangle = 
a \hat{[]}_{n,p} (b_{1} \hat{.} \cdots \hat{.} b_{p} . \llbracket
\uparrow^{n} \rrbracket)$.
\end{itemize}
and we keep the same denotation for the function symbols and the
predicate symbols.

As ${\cal N}$ is a model of the rewrite rules of $\sigma$, ${\cal M}$
is an intensional functional structure and verifies the coherence
property. Hence, it is a model of ${\cal L}$.

Then, we prove by induction over the structure of terms (resp. proposition)
that if $t$ is a term in binding logic, then the denotation
of $t$ in ${\cal M}$ is the denotation of $t'$ in ${\cal N}$. 

The only non trivial case is when $t$ is a variable. In this case, 
$t' = x[\uparrow^{n}]$, where $n$ is the sort of $t$.  
The denotation of $t$ in ${\cal N}$ is $\phi(x) \hat{[]}_{n,0}
\llbracket \uparrow^{n} \rrbracket$.
Its denotation in ${\cal M}$ is $\phi(x) \Box_{n,0} \langle \rangle =
\phi(x) \hat{[]}_{n,0} \llbracket \uparrow^{n} \rrbracket$.
 
Thus the model ${\cal M}$ is a model of $\Gamma$ and thus there is a 
proposition of $\Delta$ that is valid in ${\cal M}$. Thus there is a
proposition $\Delta'$ is valid in ${\cal N}$.} 
}

\VC{At last we conclude:}

\begin{theorem}[Soundness and Completeness]
A sequent $\Gamma \vdash \Delta$ is provable in binding logic if and
only if for each model of $\Gamma$, there is a proposition of $\Delta$
that is valid. 
\end{theorem}

\proof{From propositions \ref{a}, \ref{b} and \ref{c}.}

\begin{remark}
Intensional functional structures can be seen as intensional versions of
Henkin's models \cite{Henkin,Andrews}. Hence, we believe that our
completeness theorem could 
also be proved by translating binding logic not to predicate logic
using de Bruijn indices and explicit substitutions, but to
higher-order logic using higher-order abstract syntax.  
\end{remark}

\section*{Conclusion}

We have defined a logic with binders, that is a simple extension of
predicate logic: proof rules are the same and no extra axiom is
added. Many theories can be expressed in this framework. The
expression of a theory is close to its expression in the informal
language of mathematics.  Some theories, however, require axiom
schemes.

We have given a notion of model for this logic and proved a soundness
and completeness theorem. The first naive idea of extensional model is
not sufficient because some non provable propositions are always valid
in such models.  Our models are algebraic, or categorical, structures
that permit to speak about intensional functions both in intensional
and in extensional mathematics. We have used this notion of model to
prove that the extensionality scheme is independent.

Along the way, we have defined a translation of binding logic into
predicate logic. This translation may also be interesting for
itself. 

We conjecture that if for each binder we have an extensionality scheme
then we can restrict to models where $M_{n}$ is the set of functions
of $A^{n}$ to $A$ for some set $A$.  Other aspects of binding logic
such as unification, skolemization and proof search remain to be
studied.

\section*{Acknowledgements}

We want to thank Pierre-Louis Curien for helpful remarks on this paper.

\end{document}